\documentclass[12pt]{article}
\usepackage{epsfig}
\usepackage{amsfonts}
\usepackage{amsopn}
\usepackage{amssymb}
\usepackage{amsmath}
\usepackage{graphicx}
\usepackage{amsthm}
\usepackage{authblk}
\usepackage{bbold}
\usepackage[left=2.1cm, right=2.1cm]{geometry}

\title{Variational orthogonalization}
\author[1]{Farrokh Atai \thanks{farrokh@kth.se}}
\author[2]{Jens Hoppe}
\author[2]{ Mariusz Hynek \thanks{mkhynek@kth.se}}
\author[1]{Edwin Langmann \thanks{langmann@kth.se}}

\affil[1]{Department of Theoretical Physics, 
Royal Institute of Technology KTH, 
 106 91 Stockholm, Sweden}
\affil[2]{Department of Mathematics,
Royal Institute of Technology,
KTH, 100 44 Stockholm,
Sweden}
\begin{document}
\date{\today}
\maketitle
\thispagestyle{empty}
\abstract{We introduce variational methods for finding approximate eigenfunctions and eigenvalues of quantum Hamiltonians by constructing a set of orthogonal wave functions which approximately solve the eigenvalue equation.}

\section{Introduction}
Most models in physics have not been solved exactly and can not be treated perturbatively since their Hamiltonians do not contain any small parameter.  It is therefore  useful to introduce methods for finding an approximation to the spectrum (and corresponding eigenfunctions). Our original motivation was a class of matrix models, where a crucial role is played by an $SU(N)$ invariant Hamiltonian with quartic interaction \cite{matrixmodel}
\begin{equation}
H = -\sum_{a=1}^{N^2-1}\sum_{i=1}^d\frac{\partial^2}{\partial q_{i,a}^2}+\frac{1}{2}\sum_{a,b,c,b',c'=1}^{N^2-1}\sum_{i,j=1}^d f_{a b c}^{(N)}f_{a b' c'}^{(N)}q_{i,b}q_{i,b'} q_{j,c} q_{j,c'}
\label{mm}
\end{equation}
The spectrum of (\ref{mm}) is not known yet, neither analytically nor numerically, even for the simplest possible case $N=2$ 
\begin{equation}
H= -\sum_{i=1}^{d}\vec{\nabla}^2_i+\sum_{i,j=1}^d(\vec{q}_i \times \vec{q}_j)^2
\label{su2}
\end{equation} 
For $d=2$ the latter Hamiltonian can be reduced to ($x \geq y \geq0$) 
\cite{hoppegold} (see Appendix B for details) 
\begin{equation}
H=-\frac{\partial^2}{\partial x^2}-\frac{\partial^2}{\partial y^2}+x^2y^2-\frac{1}{4}(\frac{1}{x^2}+\frac{1}{y^2})-\frac{(x^2+y^2)}{(x^2-y^2)^2}
\label{x2y2}
\end{equation}
acting on $SU(2) \times SO(2)$ invariant wavefunctions vanishing $\sim \sqrt{xy(x^2-y^2)}$ at the singular points. The two methods presented in this paper are applicable to the above problems and computationally "cheaper"  (due to their simplicity) than the standard approximate diagonalization methods.

We first test them for two toy models:  the anharmonic oscillator
\begin{equation}
H=-\frac{\partial^2}{\partial x^2}+x^4
\label{anhosc}
\end{equation}
and a simplified version of the Hamiltonian in (\ref{x2y2})
\begin{equation}
H=-\frac{\partial^2}{\partial x^2}-\frac{\partial^2}{\partial y^2} +x^2y^2, 
\label{x2y2unred}
\end{equation}
and then present some results for the $N=2$ matrix model (\ref{su2}), $d$ arbitrary. 

\section{Variational orthogonalization}
The main idea is to construct a set of orthogonal functions, which approximately solve the Schr\"odinger equation
\begin{equation}
H \psi= E \psi
\end{equation}
for a given hermitian operator $H$ acting on a certain Hilbert space $\mathcal{H}$ (with scalar product $\langle \cdot, \cdot \rangle$ and corresponding norm $||f||^2=\langle f, f \rangle$). Assume that the Hamiltonian $H$ has a purely discrete spectrum (this is true for all Hamiltonians discussed in the introduction \cite{Simon}), and denote the symmetry group of the system by $G$. The Hilbert space splits into the direct sum of $H$- and $G$- invariant subspaces
\[
\mathcal{H}= \oplus_i  \mathcal{H}_i
\]
where the $ \mathcal{H}_i$'s are invariant subspaces of both $H$ and $G$.

Let us start with the ground state. One can introduce an ansatz for the ground state wave function  $\psi_0=\psi_{0}(\vec{x};\vec{\omega})$ (in every symmetry sector $\mathcal{H}_i$ separately), which depends on a certain number of variational parameters $(\omega^{(1)},...,\omega^{(k)})=\vec{\omega}$. Since a priori the Schr\"odinger equation is not satisfied exactly, we get
\begin{equation}
H \psi_{0}(\vec{x};\vec{\omega})= E(\vec{\omega})\psi_{0} (\vec{x};\vec{\omega})+ \chi_0(\vec{x};\vec{\omega})
  \end{equation}
An approximation of the ground state energy $E_0$ can be found by minimizing the relative norm of $\chi_0$, i.e.:
\begin{equation}
\min_{\vec{\omega},E} \frac{||\chi_0 ||^2}{||\psi_0 ||^2}=\min_{\vec{\omega},E} \frac{||(H-E)\psi_0 ||^2}{||\psi_0 ||^2}
\label{min}
\end{equation}
To extend this to excited states we introduce a variational basis of $ \mathcal{H}_i$,  $\lbrace f_0, f_1,f_2,... \rbrace$, 
consisting of normalizable functions 
\begin{equation}
f_n=f_n(\vec{x};\vec{\omega})
\end{equation}
depending on  $k$ variational parameters $\vec{\omega}$ and such that $Hf_n$ is normalizable. Then we introduce an orthogonal set of variational wave functions
\begin{equation}
\begin{array}{c}
    \psi_0(\vec{x};\vec{\omega_0})=f_0(\vec{x};\vec{\omega}_0)\\ 
    \psi_1(\vec{x};\vec{\omega}_1)= c_{10} f_0(\vec{x};\vec{\omega}_1)+f_{1}(\vec{x};\vec{\omega}_1)\\ 
    \vdots \\ 
   \psi_n(\vec{x};\vec{\omega}_n)=\sum_{l=0}^{n-1} c_{nl} f_l(\vec{x};\vec{\omega}_n) + f_n(\vec{x};\vec{\omega}_n)
\end{array}
\label{ortpol}
\end{equation}
with the orthogonality conditions
\begin{equation}
\langle \psi_i,\psi_j \rangle=0, ~~i \neq j
\label{orthcon}
\end{equation}
where $c_{nl}$ are constants. The construction of the set (\ref{ortpol}) is a recursive procedure. We start by fixing $\vec{\omega_0}$ by using (\ref{min}) and get an approximate ground state wave function $\psi_0(\vec{x})$. Then we consider the first excited state, namely $\psi_1(\vec{x};\vec{\omega_1})$. The orthogonality condition $\langle \psi_0,\psi_1 \rangle=0$ fixes the value of $c_{10}$, and then we use an analogue of (\ref{min}), i.e. we minimize 
\begin{equation}
\frac{||(H-E)\psi_1 ||^2}{||\psi_1 ||^2}
\label{min1}
\end{equation}
In general the $n$-th excited variational state is constructed by fixing the constants $c_{n,0},...,c_{n,n-1}$ using the orthogonality conditions $\langle \psi_j,\psi_n \rangle=0$ for $j=0,\ldots,n-1$ and minimizing the relative norm of $ \chi_n$:
\begin{equation}
R_n^2 :=\min_{\vec{\omega_n},E} \frac{||(H-E)\psi_n ||^2}{||\psi_n ||^2}
\label{mink}
\end{equation}
which fixes the variational parameters $\vec{\omega}_n,E$ and thus determines $E_n^{\text{approx.}}=E^{\text{min}}$ (approximate eigenvalue) and $\psi_n(\vec{x},\vec{\omega}_n^{\text{min}})$ (approximate eigenfunction).
 
 The quantities $R_n$ defined in (\ref{mink}) are a measure for  the accuracy of our approximation: as discussed in Appendix A, generically 
 \begin{equation} 
 |E_n-E_n^{\text{approx.}}|\leq  R_n
 \end{equation} 
In our test cases we found that $R_n$ is typically larger by one order of magnitude than $ |E_n-E_n^{\text{approx.}}|$. 

Note that every symmetry sector (an irreducible representation of $G$) is an invariant subspace of $H$, so in every  $\mathcal{H}_i$ the described procedure is performed independently. 

\subsection{Results for $-\partial_x^2+x^4$}
Let us present some results obtained for the anharmonic oscillator 
Hamiltonian in (\ref{anhosc}). The symmetry group of the system is $\mathbb{Z}_2$, so the Hilbert space $\mathcal{H}=L^2(\mathbb{R})$ splits into two invariant subspaces: even and odd functions of one variable. We define 
\begin{equation} 
g_n(x;\omega)= x^{n} e^{-\omega x^2/2}
\label{gn} 
\end{equation} 
and make the following choice of the variational basis basis,  $f_n=g_{2n}$ in the even sector and $f_n=g_{2n+1}$ in the odd sector ($n=0,1,2,\ldots$). The results (which turn out to be relatively accurate) are presented in Table 1. 

In order to improve them one can generalize the ansatz above to
\begin{equation} 
g_n(x;\vec{\omega})= x^{n} e^{-\omega^{(1)} x^2/2 -\omega^{(2)} x^4/4}
\label{gn2} 
\end{equation} 

Table 3 contains the results obtained for the second ansatz.

\subsection{Results for $-\partial_x^2-\partial_y^2+x^2y^2$}

The symmetry group is the point group $C_{4v}$ generated by
\begin{itemize}
\item reflection w.r.t. the $x$ axis: $(x,y)\rightarrow (-x,y)$
\item reflection w.r.t. the $y$ axis: $(x,y)\rightarrow (x,-y)$
\item reflection across the line $y=x$: $(x,y)\rightarrow (y,x)$
\end{itemize}
so the irreducible representations can be labeled by their transformation properties under the action of the three generators above (Even or Odd). There exist 5 irreducible representations of $C_{4v}$: $EEE,EEO,OOE,OOO$ (1-dimensional) and one two dimensional $EO-OE$.

Let us define the following density function
\begin{equation}
\rho(x,y;\vec{\omega})= e^{-y^2 \omega^{(1)}-x^2 \omega^{(2)}-x^2 y^2 \omega ^{(3)}}+e^{-x^2 \omega ^{(1)}-y^2 \omega ^{(2)}-x^2 y^2 \omega^{(3)}}
\end{equation}

As a set of orthogonal variational wave functions in the EEE sector we take:
\begin{equation}
\begin{array}{c}
 \psi_0(x,y;\vec{\omega_0})=\rho(x,y;\vec{\omega}_0)\\ 
    \psi_1(x,y;\vec{\omega}_1)=\rho(x,y;\vec{\omega}_1) (c_{10}+ (x^2+y^2))\\ 
    \psi_2(x,y;\vec{\omega}_2)=\rho(x,y;\vec{\omega}_2) (c_{20}+ c_{21}(x^2+y^2)+ (x^4+y^4))\\   \vdots\\
\end{array}
\label{ortpolEEE}
\end{equation}

Table 5 contains the results.

\section{Variational orthogonalization - another approach}
In this chapter we introduce a practical improvement making the method previously described less demanding computationally. Instead of the set of variational wave  functions (\ref{ortpol}) we take 
\begin{equation}
\begin{array}{c}
    \psi_0(\vec{x};\vec{\omega}_0)=f_0(\vec{x};\vec{\omega}_0)\\ 
    \psi_1(\vec{x};\vec{\omega}_1)= c_{10} \psi_0(\vec{x};\vec{\omega}_0)+f_{1}(\vec{x};\vec{\omega}_1)\\ 
    \vdots \\ 
    \psi_n(\vec{x};\vec{\omega}_n)=\sum_{l=0}^{n-1} c_{nl} \psi_l(\vec{x};\vec{\omega}_l)+f_n(\vec{x};\vec{\omega}_n)\\
\end{array}
\label{ortpol2}
\end{equation}
which makes the orthogonality conditions (\ref{orthcon}) much simpler to solve,
\begin{equation}
c_{ni}=-\frac{\langle f_n(\cdot; \vec{\omega}_n),\psi_i(\cdot; \vec{\omega}_i) \rangle}{||\psi_i(\cdot;\vec{\omega}_i)||^2}, ~0\leq i <n,
\end{equation}
and speeds up the computation. 

Tables 2 and 4 show the results for the anharmonic oscillator obtained with this method. 

One can also apply the new approach to the $x^2y^2$ model (\ref{x2y2unred}). As a set of variational wave functions in the EEE sector one can take 
\begin{equation}
\begin{array}{c}
 \psi_0(x,y;\vec{\omega_0})=\rho(x,y;\vec{\omega}_0)\\ 
    \psi_1(x,y;\vec{\omega}_1)=c_{10}\psi_0(x,y;\vec{\omega}_0)+(x^2+y^2)\rho(x,y;\vec{\omega}_1)\\ 
    \psi_2(x,y;\vec{\omega}_2)= c_{20}\psi_1(x,y;\vec{\omega}_0)+ c_{21}\psi_1(x,y;\vec{\omega}_1)+(x^4+y^4)\rho(x,y;\vec{\omega}_2)\\   \vdots\\
\end{array}
\label{ortpolEEEn}
\end{equation}
while in the EEO sector
\begin{equation}
\begin{array}{c}
 \psi_0(x,y;\vec{\omega_0})=(x^2-y^2)\rho(x,y;\vec{\omega}_0)\\ 
    \psi_1(x,y;\vec{\omega}_1)=c_{10}\psi_0(x,y;\vec{\omega}_0)+(x^4-y^4)\rho(x,y;\vec{\omega}_1)\\ 
    \psi_2(x,y;\vec{\omega}_2)= c_{20}\psi_1(x,y;\vec{\omega}_0)+ c_{21}\psi_1(x,y;\vec{\omega}_1)+(x^6-y^6)\rho(x,y;\vec{\omega}_2)\\   \vdots\\
\end{array}
\label{ortpolEEO}
\end{equation}
Table 6 and 7 shows the result.
\section{Results for the $SU(2)$ Matrix Model}
We can apply our method to the Hamiltonian given by (\ref{su2}) for the simplest case $d=2$ and find an approximation of the two first eigenfunctions and eigenvalues in the maximal symmetry sector.
\begin{eqnarray}
\psi_0(\vec{q})&=&e^{-\omega_0\sum_{i=1}^2 \sum_{a=1}^3 q_{i a}^2/2} \nonumber \\
\psi_1(\vec{q})&=&c_{10}\psi_0(\vec{q})+e^{-\omega_1\sum_{i=1}^2 \sum_{a=1}^3 q_{i a}^2/2}\sum_{i=1}^2 \sum_{a=1}^3 q_{i a}^2 
\label{ansatzsu2}
\end{eqnarray}
where
\[
c_{10}= - \frac{48 \omega_{0}^3}{\left(\omega_{1} + \omega_{0} \right)^{4}}
\]
Table 8 shows the result (obtained numerically). 
\subsection{Analytical results for $N=2$, $d$ arbitrary} 
In order to generalize the above result to arbitrary $d$, using the same type of ansatz as in (\ref{ansatzsu2})
\begin{eqnarray}
\psi_0(\vec{q})&=&e^{-\omega_0\sum_{i=1}^d \sum_{a=1}^3 q_{i a}^2/2} \nonumber \\
\psi_1(\vec{q})&=&c_{10}\psi_0(\vec{q})+e^{-\omega_1\sum_{i=1}^2 \sum_{a=1}^3 q_{i a}^2/2}\sum_{i=1}^d \sum_{a=1}^3 q_{i a}^2,
\label{ansatzsu2d}
\end{eqnarray}
where
\begin{equation}
c_{1,0} = -\frac{3d}{\omega_1 + \omega_{0}} \left( \sqrt{\frac{2\omega_{0}}{\omega_1 + \omega_{0}} }\right)^{3d},
\end{equation}
compute the error measure
\begin{equation}
R(\psi_0)=\sqrt{\frac{\langle\psi_0,H^2\psi_0\rangle}{||\psi_0||^2}-\frac{\langle\psi_0,H\psi_0\rangle^2}{||\psi_0||^2}}.
\end{equation} 
One gets 
\begin{equation}
\frac{\langle\psi_0,H^2\psi_0\rangle}{||\psi_0||^2}(\omega_0) = \frac{3}{4} d (2 + 3 d) \omega_0^{2} + \frac{3}{4} d (d-1)(3d-4)\frac{1}{\omega_0} + \frac{3}{16} d(d-1)(d+2)(3d-1) \frac{1}{\omega_0^{4}}
\nonumber
\end{equation}

\begin{equation}
\frac{\langle\psi_0,H\psi_0\rangle}{||\psi_0||^2}(\omega_0) = \frac{3}{2} d \omega_0 + \frac{3}{4} d(d-1) \frac{1}{\omega_0^{2}} \Rightarrow \left< H\right>_{0}^{2} = \frac{9}{4} d^{2} \omega_0^{2} + \frac{9}{4} d^{2}(d-1) \frac{1}{\omega_0} + \frac{9}{16} d^{2} (d-1)^{2} \frac{1}{\omega_0^{4}} \nonumber
\end{equation}

yielding
\begin{equation}
R\left(\psi_0\right)^{2}(\omega_0) = \frac{3}{2}d \omega_0^{2} - 3 d (d-1) \frac{1}{\omega_0} + \frac{3}{8} d \left(d-1\right) \left(4d-1\right) \frac{1}{\omega_0^{4}}.
\end{equation}

Finding the minimum of $R^{2}$ simplifies to the characteristic equation $2\omega_0^{6} + 2\left(d-1\right)\omega_0^{3} - 4d^{2} +5d - 1 =0$
which can easily be solved by making the substitution $z= \omega_0^{3}$ which means finding the roots of a second order polynomial $z^{2} + d\left(d-1\right) - \frac{1}{2}\left(d-1\right)\left(4d-1\right) = 0$ and taking the positive (real) solutions of $\omega_0 = z^{1/3}$, which leads to

\begin{align}
\omega_{0}^{\text{min.}}(d)= \left(\frac{1}{2}\left( 1 - d +  \sqrt{3 \left(d-1\right) \left(3d-1\right)} \right) \right)^{1/3} \nonumber \\
E_{0}^{\text{approx.}}(d) = \frac{3d \sqrt{\left(3\left(d-1\right)\left(3d-1\right)\right)}}{\left(4 \left(1-d+\sqrt{3\left(d-1\right)\left(3d-1\right)}\right)\right)^{2/3}} 
\\
R^{2}_0(d) = 18 \frac{d(d-1) \left( 6d-3-2 \sqrt{3 \left(d-1\right)\left(3d-1\right)}\right)}{\left( 4\left( 1 - d + \sqrt{3 \left(d-1\right)\left(3d-1\right)}\right)\right)^{4/3}}. \nonumber
\end{align}
The large $d$ asymptotic behaviour of the above quantities is
\begin{align}
\omega_{0}^{\text{min.}}(d) \simeq d^{1/3} \nonumber
\\
E_{0}^{\text{approx.}}(d) \simeq \frac{9}{4} d^{4/3}
\\
R^{2}_0(d) \simeq \frac{9}{4} d^{2/3}. \nonumber
\end{align}

Let us therefore consider
\begin{equation}
\tilde{H}=d^{-\frac{4}{3}}( -\sum_{i=1}^{d}\vec{\nabla}^2_i+\sum_{i,j=1}^d(\vec{q}_i \times \vec{q}_j)^2)
\label{su2d}
\end{equation}
Then the corresponding approximation of the ground state energy of $\tilde{H}$ and its error squared read
\begin{align}
E_{0}^{\text{approx.}}(d) \simeq \frac{9}{4} \nonumber
\\ 
R^{2}_0(d) \simeq \frac{9}{4} d^{-2},
\end{align}
the regularized result becoming more and more accurate when $d$ increases. \newline
We can use the observation that $\psi_0(\omega_{0}^{\text{min.}}(d))$ is a good approximation of the ground state wave function (at least for large $d$) and get an approximation of the energy of the first excited state 
\begin{equation}
E_1^{\text{approx.}}(d)=\min_{\omega_1}\frac{\langle\psi_1,\tilde{H}\psi_1\rangle}{||\psi_1||^2}(\omega_1),
\end{equation}
which means that we probe a subspace of the orthogonal complement of the approximate ground state wave function. We find that 
\begin{multline}
 \langle\psi_1,\tilde{H}\psi_1\rangle =d^{-\frac{4}{3}}\Bigl( \left(\frac{\pi}{\omega_1}\right)^{3d/2} \left( \frac{3}{8}\frac{d\left(9 d^{2} -6d + 8\right)}{\omega_1} + \frac{9}{16} \frac{d\left(d-1\right)\left(d+2\right)\left(3d+4\right)}{\omega_1^{4}} \right)\Bigr. \\ + \left(\frac{2 \sqrt{\pi \omega_{0}}}{\omega_1 + \omega_{0}}\right)^{3d} \left( \left( -\frac{81}{2} d^{3} \omega_{0} + \frac{27}{4} d^{3} \left(d-1\right) \frac{1}{\omega_{0}^{2}} \right) \frac{1}{\left( \omega_1 + \omega_{0}\right)^{2}} + \right. \\ \Bigl. \left. \frac{18d^{2}\left(3d+2\right) \omega_{0}^{2}}{\left( \omega_1 + \omega_{0}\right)^{3}} - \frac{18d^{2} \left(d-1\right) \left(3d+4\right)}{\left(\omega_1 + \omega_{0}\right)^{4}} \right) \Bigr) \nonumber
\end{multline}
and 
\begin{equation}
\left|\left| \psi_{1} \right| \right|^{2} = \left< \psi_{1} , \psi_{1} \right>\left(\omega_1\right) = \frac{3}{4}d(3d+2) \frac{1}{\omega_1^{2}} \left( \sqrt{\frac{\pi}{\omega_1}}\right)^{3d} - \frac{9d^{2}}{\left(\omega_1+\omega_{0}\right)^{2}}\left( \frac{2 \sqrt{\pi \omega_{0}}}{\omega_1 + \omega_{0}}\right)^{3d} \nonumber
\end{equation}
Table 10 shows the results for the non-rescaled Hamiltonian (\ref{su2}) for $d=2$, which are consistent with the purely numerical results (c.p. Table 8). Table 11 and 12 contain the dependence on $d$ of our results for the ground state and the first excited state respectively. 

\subsection{Cut-off results for $N=d=2$} 
\label{cutoffresults} 
There exists an independent way to check the result of the variational orthogonalization for $d=2$. To assess the quality of our approximation we also computed the eigenvalues of the $N=d=2$ matrix Hamiltonian in the maximally symmetric sector by diagonalizing the Hamiltonian in (\ref{x2y2}) using a conventional method.

The Hamiltonian in  (\ref{x2y2}) can be written as (see Appendix B)
\begin{equation}
H=-\frac{1}{r^5}\partial_r(r^5  \partial_r)+\frac{16}{r^2} \frac{1}{\sin \theta}\partial_{\theta}(\sin \theta \partial_{\theta})+\frac{r^4}{8}(1-\cos \theta)
\label{H5}
\end{equation}
on the Hilbert space with scalar product of functions $f(r,\theta)$ with integration measure $r^5dr\sin\theta d\theta$ ($r\in[0,\infty)$, $\theta\in[-\pi,\pi]$).%

%

We work with the following basis 
\[
f_{ln}(\theta,r)=\tilde{P}_l(\cos\theta)\phi_n(r)
\]
where $\tilde{P}_l=\frac{\sqrt{2}}{\sqrt{2l+1}}P_l$ are orthonormalized Legendre polynomials 
%
and 
\[
\phi_n(r)=\frac{\sqrt{n!}}{\sqrt{(n+5)!}} L^{(5)}_{n}(r)
e^{-r/2}
\] 
are orthonormal on $[0,\infty)$ w.r.t. the weight $r^5$. Then the following matrix representation of $H$
\begin{equation}
H_{ln,l'n'}= \langle f_{l'n'},H f_{ln} \rangle 
\end{equation}
is symmetric.

In order to make it a proper matrix (with two indices) we use the inverse of the pairing function $p(l,n) = (l+n)(l+n+1)/2+n$, 
\[
H_{ab}:=H_{p^{-1}(a) ,p^{-1}(b)}
\]
We introduce a cut-off parameter $N$
\begin{equation}
H^{(N)}_{ab}=H_{ab}, ~~0\leq a,b\leq N
\label{cutoffH}
\end{equation}
and end up with an $(N+1)\times (N+1) $ matrix $H^{(N)}$, which we diagonalize numerically in Mathematica getting $N$ eigenvalues which are upper bounds on the real eigenvalues (see e.g. \cite{simonbook})
\begin{equation}
 E_i \leq E_i^{(N)}, ~~i=1,2,...,N
\end{equation}
where $E_i^{(N)}$ is the $i$-th eigenvalue of $H^{(N)}$ and $ E_i $ is the $i$-th eigenvalue of $H$. Figure 1 shows the results. 

Our benchmark results for the $N=d=2$ matrix model in the maximally symmetric sector thus obtained are presented in Table 9 (the lowest upper bounds we got) and in Figure 1 (convergence of the eigenvalues of $H^{(N)}$ with increasing $N$).  



\section{Discussion}
From a conceptual point of view, the first method (section 2) is seems more natural. For the anharmonic oscillator, it gives approximate energy  eigenfunctions 
\begin{equation} 
\psi_n(x) = P_n(x) e^{-\omega_n x^2/2}  
\end{equation} 
with the $P_n$ being natural generalizations of the Hermite polynomials: they are polynomials determined (up to normalization) by the parameters $\{ \omega_m\}_{m=0}^n$, and they provide an orthogonal basis. This motivates to define and study similar generalizations of other orthogonal polynomials. The second method (section 3) is less demanding from a computational point of view, and it also seems to give more accurate results. 

The accuracy of the energy eigenvalues can be improved if one determines the ground state wave function so as to minimize the energy expectation value 
\begin{equation} 
\langle H\rangle_\psi:= \frac{\langle\psi,H\psi\rangle}{||\psi||^2}
\end{equation} 
and use (\ref{mink}) only for excited states. 

While the examples we studied were motivated by our interest in matrix models with quartic interaction, our methods can easily be applied to other systems. 

\section*{Acknowledgments}
We would like to thank Francesco Calogero, Joachim Reinhardt, Maciej Trzetrzelewski and Jacek Wosiek for helpful discussions and e-mail correspondence. This work was supported by the G\"oran Gustafsson Foundation and the Swedish Research Council (VR) under contract numbers 621-2010-3708 and 621-2010-5591.

\appendix

\section{The error measure $R$}
In this section we give a few details about the error measure 
\begin{equation} 
R(\psi,E):= \sqrt{\frac{||(H-E)\psi||^2}{||\psi||^2}}
\label{R2} 
\end{equation} 
whose minimization is a key step of our method. 

Let $\{ \psi_n\}$ and $\{ E_n\}$ be the set of eigenfunctions and the corresponding eigenvalues of a Hamiltonian $H$. Denote by $\psi$ and $E$ an approximation of the $m$-th eigenfunction and the corresponding eigenvalue of $H$. Assuming that the eigenfunctions $\psi_n$ form a complete set we can write (assuming $||\psi||=1$)
\begin{equation}
\psi=\sum_n c_n \psi_n
\end{equation}
with $\sum_n|c_n|^2=1$, which gives 
\begin{equation}
R(\psi,E)^2=(E_m-E)^2+\sum_{n \neq m} |c_n|^2\{(E_n-E)^2-(E_m-E)^2 \}
\end{equation}
We thus get, if  $\psi$ is "closer" to $\psi_m$ than to any other eigenfunction, 
\begin{equation}
R(\psi,E)^2 \geq (E_m-E)^2
\end{equation} 

In practical computations the minimization is simplified by the following fact: the minimum of $R(\psi,E)^2$ is attained for 
\begin{equation} 
E = \frac{\langle\psi,H\psi\rangle}{||\psi||^2}, 
\end{equation} 
and thus minimizing $R(\psi,E)^2$ with respect to \ $E$ and $\psi$ is equivalent to minimizing 
\begin{equation} 
\frac{||H\psi||^2}{||\psi||^2} - \left(\frac{\langle\psi,H\psi\rangle}{||\psi||^2}\right)^2
\end{equation} 
with respect to $\psi$.  

\section{$O(2)\times O(3)$ symmetry reduction}
The coordinates appearing in (\ref{x2y2unred}) can be thought of as elements of a rectangular matrix $Q=(q_{i,a})_{i=1,...,d;a=1,2,3}$ whose singular value decompositions
\begin{equation}
Q=R \Lambda S^{T}
\label{rls}
\end{equation}
with $R \in O(d)$, $S \in O(3)$ and $\Lambda$ being a $d\times 3$ matrix with positive elements $\Lambda_{ i,a}=\delta_{i,a}\lambda_i$. For $d=2$ we can write 
\begin{equation}
Q=\begin{matrix}
\left( \begin{array}{cc}
\cos \phi & -\sin \phi  \\
         \sin \phi &  \cos \phi  \end{array} \right)
         
 \left( \begin{array}{ccc}
x &  0 & 0\\
0 &  y & 0\end{array} \right) (\vec{v_1} \vec{v_2} \vec{v_3})^T, ~x \geq y \geq 0
\end{matrix}
\end{equation}
with $\vec{v}_1$, $\vec{v}_2$, and $\vec{v}_3$ being orthonormal eigenvectors of $Q^TQ$, with eigenvalues $x^2 \geq y^2 \geq 0$ (respectively).
As the integration measure $\Pi_{i,a}dq_{i,a}$ is invariant under $SO(d)$ rotations from the left ($q_{i,a} \rightarrow q'_{i,a}=T_{ij}q_{j,a}$) as well as $SO(3)$ rotations from the right ($q_{i,a} \rightarrow \tilde{q}_{i,a}=q_{i,b}\tilde{T}_{ba}$) the Jacobian $J$ for the change of variables (\ref{rls}),

\begin{equation}
\Pi_{i,a}dq_{i,a}=Jd \Lambda dR dS
\end{equation}
is independent of $R$ and $S$, hence can be calculated using $R \approx \mathbb{1}$, $S \approx \mathbb{1}$. This gives 
\begin{equation}
dQ=dR\Lambda + d\Lambda + \Lambda dS^T,
\end{equation}
with $dR$ and $dS$ antisymmetric. For $d=2$ one gets 
\begin{equation}
dQ=\left( \begin{array}{ccc}
dx &  -yd\phi-x d \theta_3 & x d \theta_2 \\
x d \phi+y d \theta_3 &  dy & -y d \theta_1 \end{array} \right)
\end{equation}
hence $J \propto xy(x^2-y^2)$ =: $\rho$, i.e. $H \psi = E \psi$ for $\psi= \psi(x,y)$ being equivalent to 
\begin{align}
(- \frac{1}{\rho} \partial_x \rho \partial_x- \frac{1}{\rho} \partial_y \rho \partial_y + x^2 y^2) \psi=  \nonumber \\
(-\partial_x^2 -\partial_y^2-(\frac{1}{x}+\frac{2x}{x^2-y^2}) \partial_x-(\frac{1}{y}+\frac{2y}{y^2-x^2}) \partial_y + x^2 y^2) \psi =E \psi 
\end{align}
then (\ref{x2y2}) follows as the effective Hamiltonian on $\tilde{\psi}:= \sqrt{\rho} \psi$ (with $\int |\tilde{\psi}|^2 dxdy=\int |\psi |^2 \rho dxdy < \infty $), while $x^2=r \cos \theta$, $y^2=r \sin \theta$ gives (\ref{H5}). 

\begin{table}[h!]
 \caption{Variational orthogonalization in the $x^n e^{-\omega_n x^2/2}$ basis  (\ref{gn}) for the anharmonic oscillator in the first approach ($E_n^{\text{approx.}}$) in comparison to the results presented in \cite{anh} ($\epsilon_n$).} 
\begin{center}
    \begin{tabular}{ | l | l | l |l |l |l | p{5cm} |}
    \hline
 n& $\epsilon_n$ &$E_n^{\text{approx.}}$ & $R_n$ &$\omega_n^{\text{min.}}$  \\ \hline
 0 &1.06036167&1.086  &  0.5  &  1.54  \\ \hline
 1 & 3.79967303&3.854  &  0.9  &  1.77\\        \hline
 2 & 7.45569794&7.536  &  1.3  &  1.96\\        \hline
 3 & 11.6447455& 11.779 &  1.7  &  2.13  \\        \hline
 4 & 16.2618261&16.430 &  2.1  &   2.16  \\        \hline
 5 & 21.2383729&21.453 &  2.6  &   2.38  \\\hline
 6 &  26.5284711&26.792 &  3.2   &  2.5 \\\hline
 7 &32.0985978& 32.414 &  3.7   &  2.6 \\\hline
 8 & 37.9230011&38.292  &  4.3 &  2.7 \\\hline
 9 & 43.9811582& 44.406  &  4.9   &   2.8\\\hline
 10 & 50.2562547& 50.739 &  5.6   &   2.8\\\hline
    \end{tabular}
\end{center}
    \end{table}

    \begin{table}[h!]
\caption{Variational orthogonalization in the $x^n e^{-\omega_n x^2/2}$ basis  (\ref{gn}) for the anharmonic oscillator in the second approach ($E_n^{\text{approx.}}$) in comparison to the results presented in \cite{anh} ($\epsilon_n$).} 
\begin{center}
    \begin{tabular}{ | l | l | l |l |l |l | p{5cm} |}
    \hline
 n& $\epsilon_n$  &$E_n^{\text{approx.}}$ & $R_n$&$ \omega_n^{\text{min.}}$  \\ \hline
 0 &1.06036167&1.086  &  0.5& 1.54   \\ \hline
 1 & 3.79967303&3.854  &  0.9 & 1.78\\        \hline
 2 & 7.45569794&7.535  &  1.3 & 1.95\\        \hline
 3 & 11.6447455& 11.767 &  1.7 & 2.10 \\        \hline
 4 & 16.2618261&16.426 &  2.1 & 2.21 \\        \hline
 5 & 21.2383729&21.448 &  2.6 & 2.35  \\\hline
 6 &  26.5284711&26.785 &  3.1 & 2.46  \\\hline
 7 &32.0985978& 32.405 &  3.7 & 2.56  \\\hline
 8 & 37.9230011&37.852  &  3.2 & 3.61\\\hline
 9 & 43.9811582& 43.900  &  3.7 & 3.56 \\\hline
 10 & 50.2562547& 50.258 &  3.9 & 3.64\\\hline
    \end{tabular}
    \end{center}
    \end{table}
    
    \begin{table}
\caption{Variational orthogonalization in the  $x^n e^{-\omega_n^{(1)}x^2/2-\omega_n^{(2)}x^4/4}$ basis (\ref{gn2}) for the anharmonic oscillator  in the first approach ($E_n^{\text{approx.}}$) in comparison to the results presented in \cite{anh}  ($\epsilon_n$).}
\begin{center} 
    \begin{tabular}{ | l | l | l |l |l |l |l | p{5cm} |}
    \hline
 n& $\epsilon_n$ &$E_n^{\text{approx.}}$ & $R_n$ &$\omega_{n}^{(1),\text{min.}}$ &$\omega_{n}^{(2),\text{min.}}$ \\ \hline
 0 &1.06036167&1.0604541 &  0.05  &  1.10&0.29  \\ \hline
 1 & 3.79967303&3.7998215 &  0.06  &  1.31&0.25\\        \hline
 2 & 7.45569794&7.4559170 &  0.08  &  1.46&0.23\\        \hline
 3 & 11.6447455&11.645054 & 0.10  &  1.59 &0.21 \\        \hline
 4 & 16.2618261&16.262261 &  0.12 &  1.70&0.20  \\        \hline
 5 & 21.2383729&21.236251 &  0.14  &   1.78&0.19  \\\hline
    \end{tabular}
\end{center}
    \end{table}

    \begin{table}[h!]
\caption{Variational orthogonalization in the $x^n e^{-\omega_n^{(1)}x^2/2-\omega_n^{(2)}x^4/4}$ basis (\ref{gn2}) for the anharmonic oscillator in the second approach ($E_n^{\text{approx.}}$) in comparison to the results presented in \cite{anh} ($\epsilon_n$).} 
\begin{center}
    \begin{tabular}{ | l | l | l |l |l |l | p{5cm} |}
    \hline
 n& $\epsilon_n$ &$E_n^{\text{approx.}}$ & $R_n$ & $\omega_{n}^{(1),\text{min.}}$& $\omega_{n}^{(2),\text{min.}}$ \\ \hline
 0 &1.06036167&1.0604541 &  0.05 &1.10& 0.292  \\ \hline
 1 & 3.79967303&3.7998215 &  0.06 &1.31&0.253 \\        \hline
 2 & 7.45569794&7.4559179 &  0.08 &1.46&0.232 \\        \hline
 3 & 11.6447455& 11.645057 &  0.10 &1.59&0.217  \\        \hline
 4 & 16.2618261&16.262244 & 0.12  &1.69&0.205  \\        \hline
 5 & 21.2383729&21.238901 &  0.14 &1.71&  0.195 \\\hline
 6 &  26.5284711&26.529044 & 0.16 &1.93&0.188   \\\hline
 7 &32.0985978& 32.102007 &  0.35 &1.95& 0.192  \\\hline
 8 & 37.9230011&37.9222  &  0.22 &2.03& 0.174 \\\hline
 9 & 43.9811582& 43.7762  &  0.48 &2.09& 0.164 \\\hline
    \end{tabular}
    \end{center}
    \end{table}

    \begin{table}[h!]
 \caption{Variational orthogonalization for the two dimensional model (\ref{x2y2unred}), in the $EEE$ sector, in the first approach ($E_n^{\text{approx.}}$) in comparison to the results presented in \cite{qchem} ($\epsilon_n$).}
 \begin{center}
    \begin{tabular}{ | l | l | l |l |l |l | l|p{5cm} |}
    \hline
 symmetry sector& $\epsilon_n$ &$E_n^{\text{approx.}}$ & $R_n$&$\omega_{n}^{(1),\text{min.}}$&$\omega_{n}^{(2),\text{min.}}$& $\omega_{n}^{(3),\text{min.}}$ \\ \hline
 $EEE_0$ &1.1082&1.1103 & 0.13&0.264&$10^{-8}$&0.142 \\ \hline
 $EEE_1$ &3.515&3.62352 &0.83&0.943&0.161&0.080\\ \hline
 $EEE_2$ &4.985& 5.05429 & 0.67 &0.157&0.736&0.073\\ \hline
 \end{tabular}
 \end{center}
 \end{table}
 
     \begin{table}[h!]
 \caption{Variational orthogonalization for the two dimensional model (\ref{x2y2unred}), in the $EEE$ sector, in the second approach ($E_n^{\text{approx.}}$) in comparison to the results presented in \cite{qchem} ($\epsilon_n$).} 
 \begin{center}
    \begin{tabular}{ | l | l | l |l |l |l | l|p{5cm} |}
    \hline
 symmetry sector& $\epsilon_n$ &$E_n^{\text{approx.}}$ & $R_n$&$\omega_{n}^{(1),\text{min.}}$&$\omega_{n}^{(2),\text{min.}}$& $\omega_{n}^{(3),\text{min.}}$ \\ \hline
 $EEE_0$ &1.1082&1.10883 & 0.09&0.385&0.190&0.126 \\ \hline
 $EEE_1$ &3.515&3.5514 &0.52 &0.172&0.917&0.069\\ \hline
 $EEE_2$ &4.985& 5.040 & 0.70 &0.164&1.062&0.056\\ \hline
 \end{tabular}
 \end{center}
 \end{table}

\begin{table}[h!]
\caption{Variational orthogonalization for the two dimensional model (\ref{x2y2unred}), in the $EEO$ sector, in the second approach ($E_n^{\text{approx.}}$) in comparison to the results presented in \cite{qchem} ($\epsilon_n$).} 
\begin{center}
    \begin{tabular}{ | l | l | l |l |l |l |l|l| p{5cm} |}
    \hline
 symmetry sector& $\epsilon_n$ &$E_n^{\text{approx.}}$ & $R_n$ &$\omega_{n}^{(1),\text{min.}}$& $\omega_{n}^{(2),\text{min.}}$&$\omega_{n}^{(3),\text{min.}}$ \\ \hline
 $EEO_0$ &3.056&3.0613& 0.14&0.187 & 0.461&0.0964\\ \hline
 $EEO_1$ &4.7528&4.76199 &0.34& 0.178&0.868 &0.0689 \\ \hline
 $EEO_2$ &6.1448&6.16628 &0.49&0.160 &1.10 &0.0563 \\ \hline
\end{tabular}
\end{center}
\end{table}

\begin{table}
\caption{Variational orthogonalization in the maximal symmetry sector of the $SU(2)$, $d=2$ Matrix Model in the second approach ($E_n^{\text{approx.}}$).} 
\begin{center}
    \begin{tabular}{ | l | l | l |l |l |l | p{5cm} |}
    \hline
n &$E_n^{\text{approx.}}$ & $R_n$&$\omega_n$  \\ \hline
 0 &4.56 &  1.3 & 1.13  \\ \hline
 1 & 9.12 &  2.7 & 1.32\\        \hline
    \end{tabular}
    \end{center}
    \end{table}

    \begin{table}
\caption{Upper bounds on the first few eigenvalues of the $SU(2)$, $d=2$ Matrix Model in the maximal symmetry sector, respectively the eigenvalues of Hamiltonian (\ref{cutoffH}) for $N=500$ ($E_n^{\text{cut-off}}$) }  
\begin{center}
    \begin{tabular}{ | l | l | l |l | p{5cm} |}
    \hline
n &$E_n^{\text{cut-off}}$   \\ \hline
 0 & 4.23  \\ \hline
 1 & 7.31  \\ \hline
 2 & 9.69  \\ \hline
 3 & 11.94  \\ \hline
 4 & 13.89  \\ \hline
    \end{tabular}
    \end{center}
    \end{table}

    \begin{table}
\caption{Variational orthogonalization in the maximal symmetry sector of the $SU(2)$, $d=2$ Matrix Model in the second approach using the analytical results from section 4.1 ($E_n^{\text{approx.}}$).} 
\begin{center}
    \begin{tabular}{ | l | l | l |l |l |l | p{5cm} |}
    \hline
n &$E_n^{\text{approx.}}$ & $R_n$&$\omega_n$  \\ \hline
 0 &4.56 &  1.13 & 1.32  \\ \hline
 1 & 9.17 & 3.33 & 1.14\\        \hline
    \end{tabular}
    \end{center}
    \end{table}
    \begin{table}
    \caption{The ground state energy approximation ($E_0^{\text{approx.}}$) of the regularized $SU(2)$ Matrix Model (\ref{su2d}) for various $d$ using the analytical results from section 4.1.} 
\begin{center}
    \begin{tabular}{ | l | l | l |l |l |l | l |l |l |l |p{5cm} |}
    \hline
  $d$                      &2    & 3     & 4   &  10  & 100  & 300 \\ \hline
 $E_0^{\text{approx.}}(d)$ &1.81 &  1.97 & 2.05 & 2.17 &  2.24& 2.25  \\ \hline
 $R_0(d)$ &0.524 &  0.352 &  0.265   & 0.106 & 0.011 &  0.004\\        \hline
\end{tabular}
    \end{center}
    \end{table}
    \begin{table}
    \caption{The first excited state energy approximation ($E_1^{\text{approx.}}$) of the regularized $SU(2)$ Matrix Model (\ref{su2d}) for various $d$ using the analytical results from section 4.1.} 
\begin{center}
    \begin{tabular}{ | l | l | l |l |l |l | l |l |l |l |l |l |l |lp{5cm} |}
    \hline
  $d$                      &2    & 3     & 4   &  10  & 100  & 300 \\ \hline
 $E_1^{\text{approx.}}(d)$ &3.64 & 3.35 & 3.14 & 2.64 &  2.29& 2.26  \\ \hline
\end{tabular}
    \end{center}
    \end{table}
    
\begin{figure}[p]
\centering
\includegraphics[width=1\textwidth]{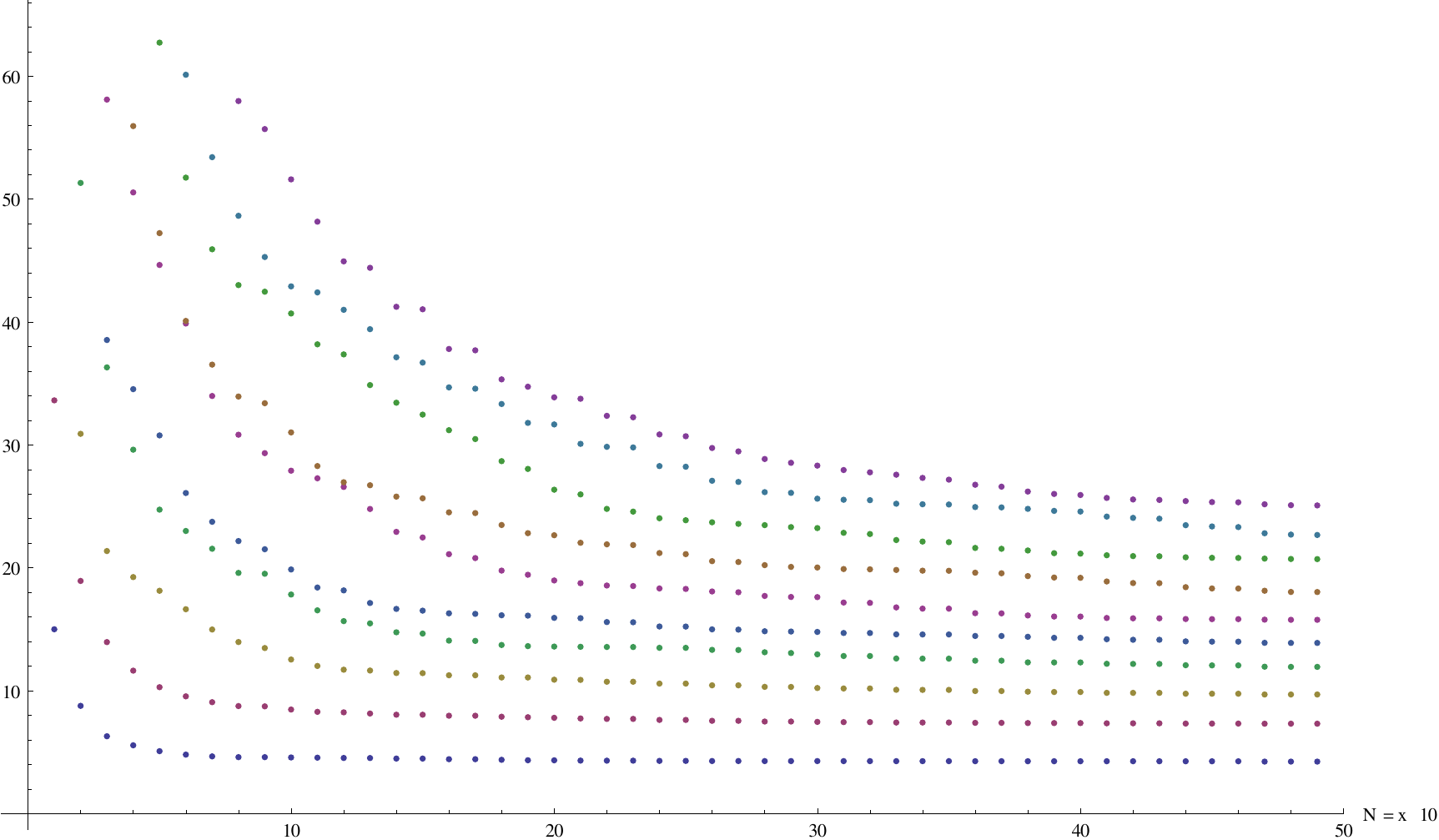}
\caption{Convergence of the eigenvalues of Hamiltonian (\ref{cutoffH}) with the increasing cut-off parameter}
\end{figure}


\clearpage
\begin{thebibliography}{7}
\bibitem{anh} F.T. Hioe, Don MacMillen and E.W. Montroll \textit{Quantum theory of anharmonic oscillators: energy levels of a single and a pair of coupled oscillators with quartic coupling}, Phys. Rept. 43 (1978) 305-335

\bibitem{qchem} Martens, Craig C.; Waterland, Robert L.; Reinhardt, William P.
 \textit{Classical, semiclassical, and quantum mechanics of a globally chaotic
system: Integrability in the adiabatic approximation}, Journal of Chemical Physics; 2/15/89, Vol. 90 Issue 4, p.2328
\bibitem{phd}
J. Hoppe, \textit{Quantum theory of a massless relativistic surface
and a two-dimensional bound state problem}, PhD Thesis MIT
1982 (http://dspace.mit.edu/handle/1721.1/15717).
\bibitem{matrixmodel}
J.Hoppe, \textit{Membranes and matrix models}, arXiv:hep-th/0206192
(IHES/P/02/47) and references therein.
\bibitem{modinv} J. Hoppe, \textit{Matrix Models and Lorentz Invariance}, J. Phys.
A 44 (2011) 055402 doi:10.1088/1751-8113/44/5/055402
arXiv:1007.5505 hep-th.

\bibitem{hoppegold} J. Goldstone, J. Hoppe, 1979, unpublished

\bibitem{Simon} B. Simon, \textit{Some quantum operators with discrete spectrum but classically continuous spectrum}, Ann. Phys. 146 (1983), 209-220

\bibitem{simonbook} M. Reed, B. Simon, \textit{Methods of modern mathematical physics, Vol. 4 Analysis of Operators}, 1978, Academic Press 



\end{thebibliography}
\end{document}